\documentclass[a4paper,10pt]{article}

\usepackage{riomacro}

\newcommand{\aap}{Astron. Astrophys. }

\newcommand{\Apj}{Astrophys.~J. }

\title{
Constraint on Heavy Element Production in 
Inhomogeneous Big-Bang Nucleosynthesis from The Light-Element Observations
}
\author{Riou Nakamura${}^{1}$, Masa-aki Hashimoto${}^{1}$, Shin-ichiro
Fujimoto${}^{2}$,Katsuhiko Sato${}^{3,4}$, \\
${}^{1}$ Department of Physics, Graduate school of sciences, Kyushu
University, \\
6-10-1 Hakozaki, Higashi-ku, Fukuoka 812-8581, Japan \\
${}^{2}$ Department of Control and Information Systems Engineering,\\
Kumamoto National College of Technology, \\
 2659-2, Suya, Koshi-Shi Kumamoto 861-1102, Japan, \\
${}^{3}$ Institute for the Physics and Mathematics of the Universe,\\
 University of Tokyo, Kashiwa, Chiba 277-8568, Japan \\
${}^{4}$ National Institutes of Natural Sciences, \\ Kamiyacho Central
Place 2F, 4-3-13 Toranomon, Minato-ku, Tokyo, 104-0001, Japan
}

\begin{document}

\maketitle

\begin{abstract}
We investigate the observational constraints on the inhomogeneous big-bang nucleosynthesis
that Matsuura et al~\cite{Matsuura2005} suggested the possibility of the
 heavy element production beyond ${}^7$Li
in the early universe.  From the observational constraints on
light elements of ${}^4_{}$He and D, possible regions  are
found on the plane of the volume fraction of the high density region
against the ratio between high- and low-density regions. 
In these allowed regions, we have confirmed that the heavy elements
beyond Ni can be produced appreciably, where $p$- and/or $r$-process
elements are produced well simultaneously.
\end{abstract}

\section{INTRODUCTION}
\label{sec:ibbn_intro}

Big-bang nucleosynthesis (BBN) has been investigated to explain the origin of the light elements, 
such as ${}^4$He, D, ${}^3$He, and ${}^7$Li, during the first few minutes~\cite{PDG2012,Iocco:2008va,Coc2012}. Standard model of BBN (SBBN) can succeed to explain the observation of those elements, 
~${}^4$He~\cite{Luridiana2003,OliveSkillman04,Izotov2010,Aver2012}, 
D~\cite{Kirkman2003,OMeara2006,Pettini2008,Pettini2012}, and
${}^3$He~\cite{Bania2002,2003ApJ...585..611V}, 
except for ${}^7$Li.
The study of SBBN has been done under the assumption of the homogeneous
universe, where the model has only one
parameter, the baryon-to-photon ratio $\eta$.
If the present value of $\eta$ is
determined, SBBN can be calculated from the thermodynamical history with use of the nuclear reaction network. 
We can obtain the reasonable value of $\eta$ by comparing the calculated abundances 
with observations. 
In the meanwhile, the value of $\eta$ is obtained as 
$\eta=\left( 5.1- 6.5 \right)\times 10^{-10}$~\cite{PDG2012}
from the observations of ${}^4$He and D. This values  agrees well with the observation of
the cosmic microwave background:~$\eta = \left( 6.19\pm 0.14 \right)\times 10^{-10}_{}$~\cite{WMAP5}.

On the other hand, BBN with the inhomogeneous baryon distribution 
also has been investigated. The model is called as inhomogeneous BBN~(IBBN).
IBBN relies on the inhomogeneity of baryon concentrations that could
be induced by baryogenesis~(e.g. Ref.~\cite{Matsuura:2004ss}) or 
phase transitions such as QCD or electro-weak phase
transition~\cite{Alcock1987,Fuller1988,IBBN_QCD} during the expansion of the
universe. Although a large scale inhomogeneity is inhibited by many 
observations~\cite{WMAP5,WMAP3}, small scale one has been advocated within the present
accuracy of the observations. Therefore, it remains a possibility for
IBBN to occur in some degree during the early era. 
In IBBN, the heavy element nucleosynthesis
beyond the mass number $A=8$ has been proposed~\cite{Matsuura:2004ss,Alcock1987,IBBN0,IBBN1,TerasawaSato89,2zone,Jedamzik1994,Wagoner1967,Wagoner1972}.
In addition, peculiar observations of abundances for
heavy elements and/or ${}^{4}_{}$He could be understood in the way of IBBN.
For example, the quasar metallicity of C, N, and Si could have been explained
from IBBN~\cite{Juarez2009}.
Furthermore, from recent observations of globular clusters, possibility
of inhomogeneous helium distribution is pointed out~\cite{Moriya2010},
where some separate groups of different main sequences in blue band of low mass stars are
assumed due to high primordial helium abundances compared to the
standard value \cite{Bedin2004,Piotto2007}.
Although baryogenesis could be the origin of the inhomogeneity, the mechanism of it
has not been clarified due to unknown properties of the supersymmetric
Grand Unified Theory~\cite{affl-dine}.

%----
Despite a negative opinion against IBBN due to insufficient consideration
of the scale of the inhomogeneity~\cite{Rauther2006},
Matsuura et al. have found that the heavy element synthesis for both $p$- and 
$r$-processes is possible if $\eta > 10^{-4}$ \cite{Matsuura2005}, where
they have also shown that the high
$\eta$ regions are compatible with the observations of the light elements,
$^4$He and D~\cite{Matsuura2007}.
However, their analysis is only limited to a parameter of
a specific baryon number concentration. 
In this paper, we extend the investigations of Matsuura et al.~\cite{Matsuura2005,Matsuura2007}
to check the validity of their conclusion from a wide parameter space of the IBBN model.

In \S \ref{sec:Numerical}, we review and give the adopted model of
IBBN which is the same one as that of  
Matsuura et al.~\cite{Matsuura2007}. Constraints on the critical parameters of IBBN
due to light element observations are shown in \S III,  and the
possible heavy element nucleosynthesis are presented in \S IV.
Finally, \S \ref{sec:summary} are devoted to the summary and discussion.

\section{Model}
\label{sec:Numerical}

In this section, we introduce the model of IBBN.
We adopt the two-zone model for the inhomogeneous BBN.
In IBBN model, we assume the existence of spherical high-density region
inside the horizon.
For simplicity, we ignore in the present study the diffusion effects before 
$\left( 10^{10}_{} {\rm K} < T < 10^{11}_{} {\rm K} \right)$ and
during the primordial nucleosynthesis $\left( 10^{7}_{} {\rm K} < T < 10^{10}_{} {\rm K} \right)$,
because the timescale of the neutron diffusion is longer than that of the cosmic expansion~\cite{IBBN0,Moriya2010}.

To find the parameters compatible with the observations, we consider the averaged
abundances between the high- and low-density regions. 
We get at least parameters for the
extreme case by averaging the abundances in two regions.
Let us  define the notations,  $n^{}_{ave}, n^{}_{high}$, and $n^{}_{low}$ as
averaged-, high-, and low- baryon number densities.
 $f^{}_v$ is the volume
fraction of the high baryon density region. 
$X^{ave}_{i}, X^{high}_i$ and $X^{low}_{i}$
are mass fractions of each element $i$ in averaged-, high- and
low-density regions, respectively, 
Then, basic relations are written as follows~\cite{Matsuura2007}: 
\begin{eqnarray}
n^{}_{ave} &=& f^{}_{v}n^{}_{high}+\left( 1-f^{}_v \right)n^{}_{low},
 \label{eq:num_b} \\
 n^{}_{ave}X^{ave}_{i} &=&  f^{}_{v}n^{}_{high} X^{high}_i 
  +\left( 1-f^{}_v \right)n^{}_{low} X^{low}_{i}. \label{eq:massfrac}
\end{eqnarray}
Here we assume the baryon fluctuation to be isothermal as was done in
previous studies~(e.g., Refs.~\cite{Alcock1987,Fuller1988,TerasawaSato89}).
Under that assumption, since the baryon-to-photon ratio is
defined by the number density of photon in standard BBN, 
Eqs.~\eqref{eq:num_b} and \eqref{eq:massfrac} are rewritten as follows:
\begin{eqnarray}
\eta^{}_{ave}  &=&
 f^{}_{v}\eta^{}_{high}+(1-f^{}_v)\eta^{}_{low},  \label{eq:eta_ave}
\\
\eta^{}_{ave}X^{ave}_{i}
 &=&
 f^{}_{v}X^{high}_{i}\eta^{}_{high}+(1-f^{}_{v})X^{low}_{i}\eta^{}_{low},
\label{eq:Yi_ave}
\end{eqnarray}
where $\eta$s with subscripts are the baryon-to-photon ratios in each
region. 
In the present paper, we fix $\eta^{}_{ave}=6.19\times10^{-10}$
%\footnote{Although this value is not the best-fit value
%$6.19\times10^{-10}$~\cite{WMAP5} for the recently data, it is within
%$1\sigma$ confidence level. Then there is no difference if we adopt the new result.}
 from the cosmic
microwave background observation~\cite{WMAP5}.
The values of $\eta^{}_{high}$ and $\eta^{}_{low}$ are obtained from both $f^{}_v$ and
the density ratio between high- and low-density region:
 $R\equiv n^{}_{high}/n^{}_{low}=\eta^{}_{high}/\eta^{}_{low}$.

To calculate the evolution of the universe, 
we solve the following Friedmann equation,
\begin{equation}
\left(\frac{\dot{x}}{x}\right)^2 =\frac{8\pi G}{3}\rho , 
\label{eq:friedman_eq}
\end{equation}
where $x$ is the cosmic scale factor and $G$ is the gravitational constant.
The total energy  density $\rho$ in Eq.~\eqref{eq:friedman_eq}  is the sum of decomposed parts:
\begin{equation}
 \rho=\rho^{}_{\gamma}+\rho^{}_{\nu}+\rho^{}_{e^{\pm}} +\rho^{}_{b}.
\label{eq:rho_total}
\end{equation}
Here the subscripts $\gamma, \nu$, and $e^{\pm}_{}$ indicate 
photons,  neutrino, and electrons/positrons, respectively.
The final term is the baryon density obtained as $\rho_{b}\simeq m^{}_pn^{}_{ave}$.

We should note about the energy density of baryon.
To get the time evolution of the baryon density in both regions, the energy conservation law is used:
\begin{equation}
\frac{d}{dt}(\rho x^3) + 
p\frac{d}{dt}(x^3)
=0, \label{eq:rho_dot}
\end{equation}
where $p$ is the pressure of the fluid. When we solve
Eq.~\eqref{eq:rho_dot}, initial values in both regions are obtained 
from Eq.~\eqref{eq:eta_ave} with $f^{}_v$ and $R$ fixed.
For $\eta^{}_{high}\geq 2\times10^{-4}_{}$,
the baryon density in the high-density region, $\rho^{}_{high}$, is larger than 
the radiation component at $T > 10^{9}$~K.
However, %the term $f^{}_v\rho^{}_{high}$ contributes to the cosmic expansion at the early epoch. 
we note that the contribute to eq.~\eqref{eq:rho_total} is not $\rho^{}_{high}$, but $f^{}_v\rho^{}_{high}$. 
In our research, the ratio of $f^{}_{v}\rho^{}_{high}$ to $\rho^{}_{\gamma}$ is about $10^{-7}_{}$ at BBN epoch.
Therefore, we can neglect the final term of eq.~\eqref{eq:rho_total} in the same way as has been done in SBBN during the calculation of eq.\eqref{eq:friedman_eq}.
%Since we can say that the high-density region has minor effect on the
%cosmic evolution, we calculate the temperature
%evolution using Eq.~\eqref{eq:friedman_eq} for both regions.

\section{Constraints from light-element observations}
\label{sec:Results}

In this section, we calculate the nucleosynthesis in high- and
low-density regions with use of the BBN code~\cite{Hashimoto1985} which
includes 24 nuclei from neutron to ${}^{16}_{}$O.
We adopt the reaction rates of Descouvemont et al.~\cite{DAA04}, the neutron lifetime 
$\tau^{}_N = 885.7$ sec~\cite{PDG2012}, and consider three massless neutrinos.

Let us consider the range of $f^{}_v$.
For $f^{}_v\ll0.1$, the heavier elements can be synthesized in the
high-density regions as discussed in Ref.~\cite{Jedamzik1994}.
For $f^{}_v>0.1$, contribution of the low-density
region to $\eta^{}_{ave}$ can be neglected and therefore
to be consistent with observations of light elements, we need to
impose the condition of $f^{}_v<0.1$.

\begin{figure}[tb]
\begin{center}
 \includegraphics[width=0.8\linewidth,keepaspectratio]{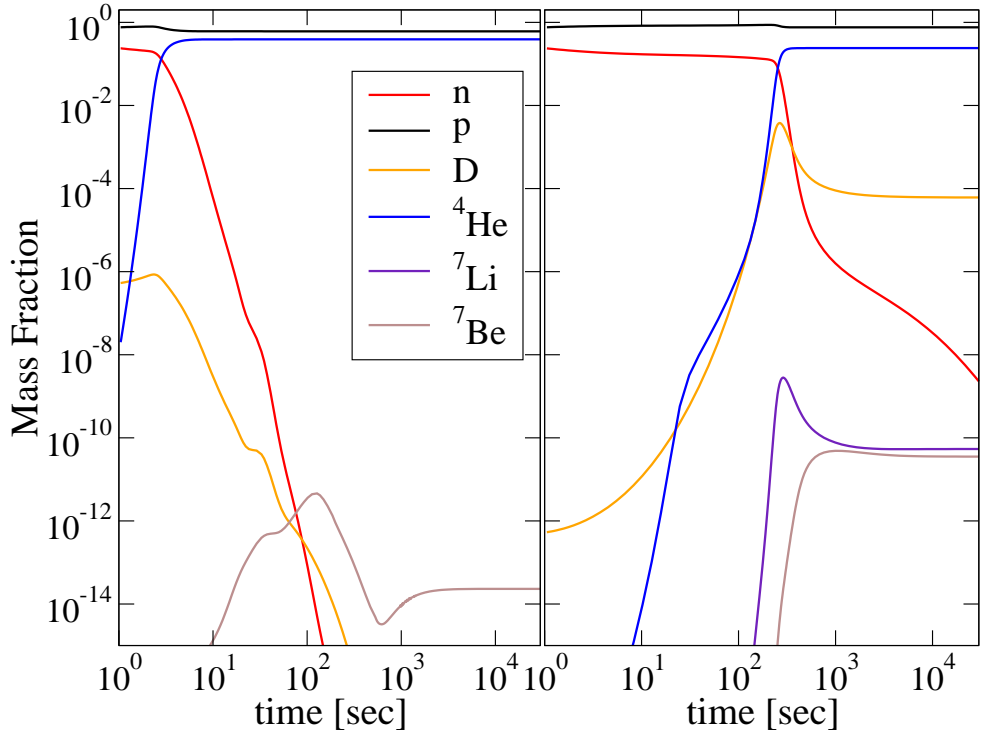}
 \caption{Illustration of the nucleosynthesis in the two-zone
 IBBN model with $f^{}_v=10^{-6}_{}$ and $R=10^{6}_{}$. 
 The baryon-to-photon ratios in the high- (left panel) and low- (right
 panel) density regions are  $\eta^{}_{high}=3.05\times10^{-4}$ and
 $\eta^{}_{low}=3.05\times 10^{-10}$,  respectively.}
\label{fig:2zone_bbn}
\end{center}
\end{figure}

Figure~\ref{fig:2zone_bbn} illustrates the light element synthesis
in the high- and low-density regions with $f_v=10^{-6}_{}$ and
$R=10^6_{}$ that corresponds to $\eta^{}_{high}=3.05\times10^{-4}$ and
$\eta^{}_{low}=3.05\times10^{-10}$.
Light elements synthesized in these calculations are shown in Table~\ref{tab:fig1_result}.
In the low-density region the evolution of the elements is almost the same 
as the case of SBBN. In the high-density region, 
while ${}^4_{}$He  is more abundant than that in the low-density region,
${}^7_{}$Li (or ${}^7_{}$Be) is much less produced.
In this case, we can see that average values such as
${}^4_{}$He and D are overproduced as shown in Table~\ref{tab:fig1_result}.
However, this overproduction can be saved by choosing the parameters carefully;
We need to find the reasonable parameter ranges for both $f^{}_v$ and $R$
by comparing with the observation of the light elements.

\begin{table}
\caption{\label{tab:fig1_result} 
The numerical abundances of light elements synthesized as shown in Figure~\ref{fig:2zone_bbn}.
  }
\begin{center}
\begin{tabular}{clll}
\hline\hline
Elements & $X^{high}_{i}$ & $X^{low}_{i}$  & $X^{}_{i}$ \\
\hline
   p &    $0.608$  &   $0.759$    &   $0.684$ \\
   D &    $3.07\times10^{-18}_{}$ &   $1.19\times10^{-4}_{}$ &  $5.95\times10^{-5}_{}$ \\
 T +${}^3_{}$He & $1.15\times10^{-13}_{}$ & $3.41\times10^{-5}_{}$ &  $1.71\times10^{-5}_{}$ \\
 ${}^4_{}$He &    $0.392$  &  $0.241$     &   $0.316$ \\
 ${}^7_{}$Li + ${}^7_{}$Be &  $8.2\times10^{-13}_{}$ &   $6.29\times10^{-10}_{}$ &  $3.14\times10^{-10}_{}$ \\
\hline\hline
\end{tabular}
\end{center}
\end{table}

Now, we put constraints on $f^{}_v$ and $R$ by comparing the average
values of ${}^4_{}$He and D obtained from Eq.~(\ref{eq:Yi_ave}) with 
the following observational values. 
First we consider the primordial ${}^4_{}$He abundance reported in
Ref.\cite{Izotov2010}:
\begin{equation}
 Y_p=0.2565 \pm 0.0010 \pm 0.0050, \notag
\end{equation}
and Ref.\cite{Aver2012}:
\begin{equation}
 Y^{}_p = 0.2534 \pm 0.0083. \notag
\end{equation}
We adopt ${}^4$He abundances as follows:
\begin{equation}
 0.2415 < Y_p < 0.2617. \label{eq:Heobs}
\end{equation}

Next, we take the primordial abundance  from the
 D/H observation reported in Ref.~\cite{Pettini2008}:
\begin{equation}
\text{D/H}  =\left( 2.84\pm0.26 \right)\times 10^{-5},\notag %\label{eq:Dobs}
\end{equation}
and Ref.~\cite{Pettini2012}:
\begin{eqnarray}
 \text{D/H} & = & \left( 2.535 \pm 0.05 \right) \times 10^{-5}, \notag \\
 \text{D/H} & = & \left( 2.48 \pm 0.12\right) \times 10^{-5}. \notag
\end{eqnarray}
Considering those observations with errors, we adopt the primordial D/H abundance 
as follows:
\begin{equation}
 2.36 < \text{D/H}\times 10^{5} < 3.02. \label{eq:Dobs}
\end{equation}

\begin{figure}[tb]
\begin{center}
 \includegraphics[width=0.8\linewidth,keepaspectratio]{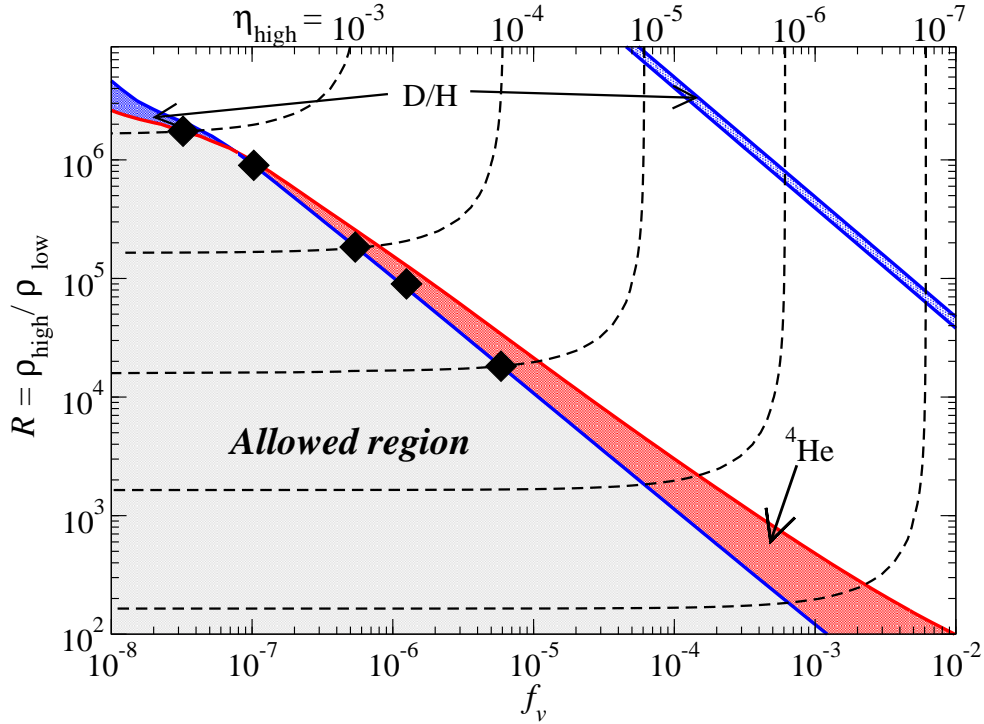}
\caption{Constraints on the $f_v-R$ plane from the observations of light
 element abundances. The region below the red line is allowable one obtained
 from ${}^4_{}$He observation \eqref{eq:Heobs}. 
 Constraints from the D/H observation \eqref{eq:Dobs} are shown by the
 region below the blue line. The gray region corresponds to the allowable 
parameters determined from the two observations of ${}^4_{}$He  and D/H. 
There is another region which is still consistent  with only D/H in the upper right direction.
This is the contribution of the low density region with $\eta^{}_{low}\sim 10^{-12}$; The D abundance tends to decrease against the baryon density for $\eta>10^{-12}$.  
The dotted lines show 
 the contours  of the baryon-to-photon ratio in the high-density
 region. Filled squares indicate the parameters for heavy element
 nucleosynthesis adopted in \S \ref{sec:Result2}.}
\label{fig:HeDcntr}
\end{center}
\end{figure}

Figure~\ref{fig:HeDcntr} illustrates the constraints on the $f^{}_v-R$ plane
from the above light-element observations with contours of constant $\eta^{}_{high}$.
The solid and dashed lines indicate the upper limits from Eqs.~(\ref{eq:Heobs})
and (\ref{eq:Dobs}), respectively.
As the results, we can obtain approximately the following relations between $f^{}_v$ and $R$ :
\begin{equation}
 R \leq 
\begin{cases}
10^4\times f_v^{-0.3} & \text{~for~} f_v>7.4\times10^{-6}, \\
0.13\times f_v^{-0.98} & \text{~for~} f_v\leq 7.4\times10^{-6} .
\end{cases}
\label{eq:HeDlimit_} 
\end{equation}
The ${}^4$He observation \eqref{eq:Heobs} gives the upper bound for $f_v<7.4\times10^{-6}$, 
and the limit for $f_v>7.4\times10^{-6}$ is obtained from D observation \eqref{eq:Dobs}.
As shown in Figure~\ref{fig:HeDcntr}, we can find the allowed regions
which include 
the very high-density region such as $\eta^{}_{high}=10^{-3}$. 

\begin{figure}[tb]
\begin{center}
  \includegraphics[width=0.80\linewidth,keepaspectratio]{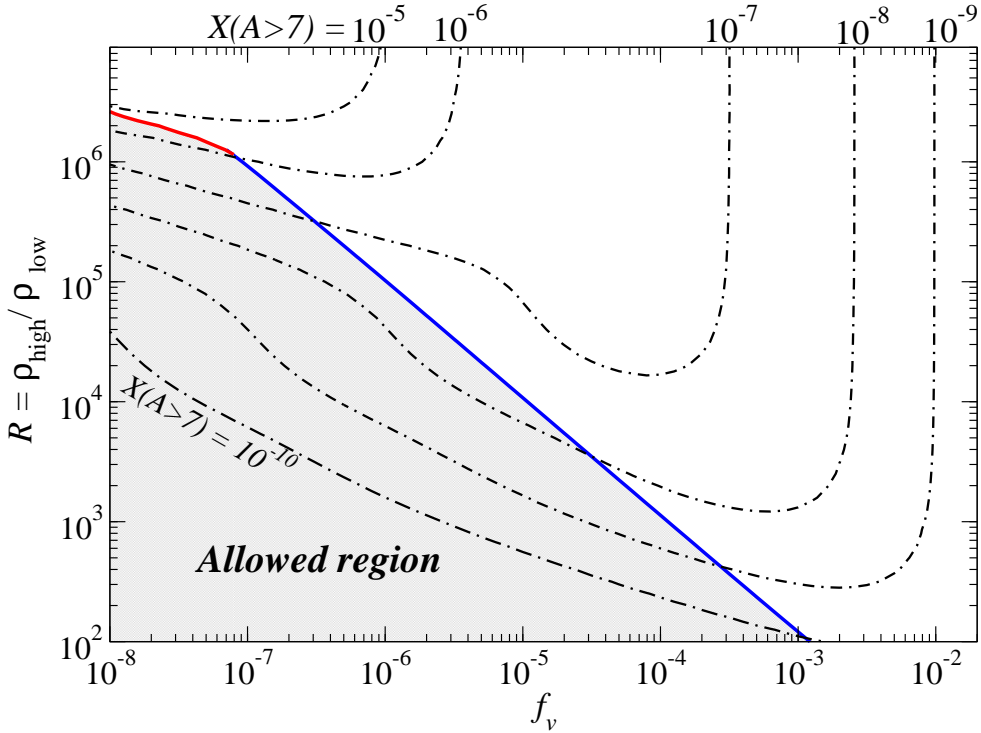}
\caption{Contours of the averaged total mass fractions which are the sum
 of nuclei heavier than  ${}^7_{}$Li, where we find consistent region of the produced elements with ${}^4_{}$He
 and D observations.}
\label{fig:cntr_upLi7}
\end{center}
\end{figure}

We should note that $\eta^{}_{high}$ takes larger value, nuclei which are heavier
than ${}^7_{}$Li are synthesized  more and more. Then we can estimate
the amount of total CNO elements in the allowed region.
Figure~\ref{fig:cntr_upLi7} illustrates the contours of the summation of
the average values of the heavier nuclei~($A>7$), which correspond to
Figure~\ref{fig:HeDcntr} and are drawn using the constraint from
${}^4_{}$He and D/H observations .
As a consequence, we get the upper limit of total mass 
fractions for heavier nuclei as follows: 
\[
X(A>7) \leq 10^{-5}_{}. 
\]

\section{Heavy element Production}
\label{sec:Result2}

In the previous section, we have obtained the amount of CNO elements produced
in the two-zone IBBN model.
However, it is not enough to examine the nuclear production beyond $A>8$ because the
baryon density in the high-density region becomes so high that elements
beyond CNO isotopes can be produced~\cite{Matsuura2005,Matsuura:2004ss,2zone,Wagoner1967}.
In this section, we investigate the heavy element nucleosynthesis in the
high-density region considering the constraints shown in Figure~\ref{fig:HeDcntr}.
Abundance change is calculated with a large nuclear reaction
network, which includes 4463 nuclei from neutron $(n)$, proton $(p)$ to Americium 
({\it Z} = 95 and {\it A} = 292).
Nuclear data, such as reaction rates, nuclear masses, and partition
functions, are the same as used in~\cite{fujimoto} except for the
neutron-proton interaction; We use the weak interaction 
of Kawano code~\cite{Kawano}, which is adequate for the high temperature
epoch of $T>10^{10}$~K.

As seen in Figure~\ref{fig:cntr_upLi7}, heavy elements of
$X(A>7)>10^{-9}$ are produced nearly along the upper limit of $R$.
Therefore, to examine the efficiency of the heavy element production, 
we select five models with the following parameters:
$\eta^{}_{high}=10^{-3}_{}, 5.1\times10^{-4}, 10^{-4}_{}, 5.0\times10^{-5}$
, and $10^{-5}$ corresponded to
$\left( f^{}_v, R \right)= \left( 3.24\times10^{-8}, 1.74\times10^{6}_{} \right)$
, $\left( 1.03\times10^{-8}, 9.00\times10^{5}_{} \right)$
, $\left( 5.41\times10^{-7}, 1.84\times10^{5}_{} \right)$
, $\left( 1.50\times10^{-6}, 9.20\times10^{4}_{} \right)$
, and $\left( 5.87\times10^{-6}, 1.82\times10^{4}_{} \right)$.
Adopted parameters are indicated by filled squares in Figure~\ref{fig:HeDcntr}.

First, we evaluate the validity of the nucleosynthesis code with $4463$ nuclei.  
Table~\ref{tab:abundance_4463} shows the results of the light elements, p, D, ${}^4_{}$He, 
${}^3_{}$He, and ${}^7$Li. The results of the high-density region is calculated by 
the extended nucleosynthesis code, and the abundances in the low-density region is obtained by 
BBN code.The averaged abundances is obtained by Eq.~\eqref{eq:Yi_ave}.  
Since the averaged values of ${}^4$He and D are consistent with the observations, 
there is no difference between BBN code and the extended nucleosynthesis code 
in regard to the averaged abundances of light elements.

\begin{table*}[tb]
\caption{\label{tab:abundance_4463} 
Mass fractions of light elements for the four cases :
 $\eta^{}_{high}\simeq10^{-3}_{}$, $\eta_{high}=5\times10^{-4}$, 
 $\eta^{}_{high}\simeq10^{-4}$, and $\eta^{}_{high}=10^{-5}$. $t^{}_{fin}$ and $T^{}_{fin}$ is the time
 and temperature at the final stage of the calculations.}
\begin{center}
{%\small
\begin{tabular}{ccccccc}
\hline\hline
$f^{}_v, R $
& \multicolumn{3}{c}{$3.23\times10^{-8}_{}, 1.74\times10^{6}_{}$}
& \multicolumn{3}{c}{$1.03\times10^{-7}_{}, 9.00\times10^{5}_{}$}
%& \multicolumn{3}{c}{$1.0\times10^{-7}_{}, 1.7\times10^{5}_{}$}
 \\
%\hline 
$( \eta^{}_{high}, \eta^{}_{low} )$  
& \multicolumn{3}{c}{($1.02\times10^{-3}_{}$, $5.86\times10^{-10}_{}$) }
& \multicolumn{3}{c}{($5.10\times10^{-4}_{}$, $5.67\times10^{-10}_{}$) }
%& \multicolumn{3}{c}{$1.02\times10^{-4}_{}$, $6.00\times10^{-10}_{}$}
 \\
\hline
$\left( t^{}_{fin}, T^{}_{fin} \right)$ & 
\multicolumn{3}{c}{$1.0\times10^{5}_{}$sec,~$4.2\times10^{7}_{}$ K} &
\multicolumn{3}{c}{$1.1\times10^{5}_{}$sec,~$4.9\times10^{7}_{}$ K} 
%\multicolumn{3}{c}{$1.2\times10^{5}_{}$sec,~$4.3\times10^{7}_{}$ K} 
\\
\hline%\hline
elements & high & low & average & high & low & average % &  high & low & average 
\\
\hline 
p   & $0.586$ & $0.753$ & $0.744$ & $0.600$  &  $0.753$ &  $0.740$  \\
D     & $1.76\times10^{-21}_{}$ & $4.50\times10^{-5}_{}$ &  $4.26\times10^{-5}_{}$ 
       & $3.43\times10^{-21}_{}$ & $4.75\times10^{-5}_{}$ &  $4.34\times10^{-5}_{}$ 
%      & $6.84\times10^{-22}_{}$ & $4.34\times10^{-5}_{}$ &  $4.27\times10^{-5}_{}$
\\
${}^3_{}$He+T &   $2.91\times10^{-14}_{}$ &  $2.18\times10^{-5}_{}$ & $2.07\times10^{-5}_{}$ 
	     &   $2.77\times10^{-14}_{}$ &  $2.23\times10^{-5}_{}$ & $2.04\times10^{-5}_{}$ \\
${}^4_{}$He 
& $0.413$ & $0.247$ & $0.256$
& $0.400$ & $0.247$ & $0.260$ 
\\
%& $0.362$ & $0.248$ & $0.249$ \\
${}^7_{}$Li+${}^7_{}$Be &
  $1.63\times10^{-13}_{}$ &  $1.78\times10^{-9}_{}$ & $1.68\times10^{-9}_{}$ 
& $6.80\times10^{-14}_{}$ &  $1.65\times10^{-9}_{}$ & $1.52\times10^{-9}_{}$ 
%&   $7.42\times10^{-13}_{}$ &  $1.87\times10^{-9}_{}$ & $1.70\times10^{-9}_{}$ 
\\
\hline\hline
\end{tabular}
{(a) For cases of $\eta^{}_{high}=10^{-3}$ and $\eta^{}_{high}=5\times10^{-4}_{}$.}

\vspace{0.5cm}

% table I(b)
\begin{tabular}{ccccccc}
\hline\hline
$f^{}_v, R $ 
%& \multicolumn{3}{c}{$2.1\times10^{-8}_{}, 1.8\times10^{6}_{}$}
%& \multicolumn{3}{c}{$7.0\times10^{-8}_{}, 9.3\times10^{5}_{}$}
& \multicolumn{3}{c}{$5.41\times10^{-7}_{}, 1.84\times10^{5}_{}$}
& \multicolumn{3}{c}{$5.87\times10^{-6}_{}, 1.82\times10^{4}_{}$}
 \\
%\hline 
$(\eta^{}_{high}, \eta^{}_{low} )$  
%& \multicolumn{3}{c}{$1.06\times10^{-3}_{}$, $5.88\times10^{-10}_{}$ }
%& \multicolumn{3}{c}{$5.33\times10^{-4}_{}$, $5.73\times10^{-10}_{}$ }
& \multicolumn{3}{c}{($1.04\times10^{-4}_{}$, $5.62\times10^{-10}_{}$)}
& \multicolumn{3}{c}{($1.02\times10^{-5}_{}$, $5.59\times10^{-10}_{}$)}
 \\
\hline
$\left( t^{}_{fin}, T^{}_{fin} \right)$ & 
%\multicolumn{3}{c}{$1.0\times10^{5}_{}$sec,~$4.2\times10^{7}_{}$ K} &
%\multicolumn{3}{c}{$1.1\times10^{5}_{}$sec,~$4.9\times10^{7}_{}$ K} &
\multicolumn{3}{c}{$1.2\times10^{5}_{}$sec,~$4.3\times10^{7}_{}$ K} &
\multicolumn{3}{c}{$1.2\times10^{5}_{}$sec,~$4.5\times10^{7}_{}$ K} 
\\
\hline%\hline
elements & high & low & average & high & low & average% &  high & low & average 
\\
\hline 
p   & $0.638$  &  $0.753$ &  $0.742$ &  $0.670$ &   $0.753$ &   $0.745$ \\
D %    & $1.76\times10^{-21}_{}$ & $4.48\times10^{-5}_{}$ &  $4.32\times10^{-5}_{}$ 
  %    & $4.14\times10^{-21}_{}$ & $4.67\times10^{-5}_{}$ &  $4.38\times10^{-5}_{}$ 
      & $6.84\times10^{-22}_{}$ & $4.79\times10^{-5}_{}$ &  $4.36\times10^{-5}_{}$
      & $1.12\times10^{-22}_{}$ & $4.48\times10^{-5}_{}$ &  $4.37\times10^{-5}_{}$
\\
${}^3_{}$He+T &   $1.63\times10^{-13}_{}$ &  $2.23\times10^{-5}_{}$ & $2.04\times10^{-5}_{}$ 
	     &  $1.49\times10^{-9}_{}$ &  $2.25\times10^{-5}_{}$ & $2.03\times10^{-5}_{}$ \\
${}^4_{}$He 
%& $0.413$ &   $0.247$  &   $0.253$
%& $0.402$ & $0.247$ & $0.257$   
& $0.362$ & $0.247$ & $0.258$ 
& $0.330$ & $0.247$ & $0.254$
\\
${}^7_{}$Li+${}^7_{}$Be %&
%  $1.63\times10^{-13}_{}$ &  $1.79\times10^{-9}_{}$ & $1.72\times10^{-9}_{}$ 
%&  $3.43\times10^{-13}_{}$ &  $1.70\times10^{-9}_{}$ & $1.59\times10^{-9}_{}$ 
& $7.42\times10^{-13}_{}$ &  $1.64\times10^{-9}_{}$ & $1.49\times10^{-9}_{}$ 
& $6.73\times10^{-8}_{}$  &  $1.62\times10^{-9}_{}$ & $7.96\times10^{-9}_{}$ 
\\
\hline\hline
\end{tabular}
{(b) For cases of $\eta^{}_{high}=10^{-4}$ and $\eta^{}_{high}=10^{-5}_{}$.}
}
\end{center}

\end{table*}

\begin{figure}[thb]
\begin{center}
 \includegraphics[width=0.6\linewidth,keepaspectratio]{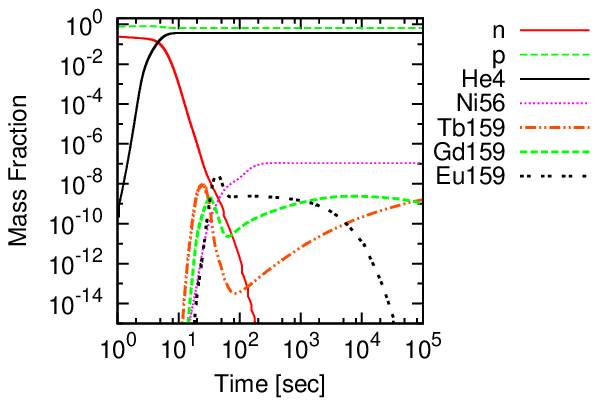}

 {(a) $\eta^{}_{high}=1.02\times10^{-4}_{}$}

 \includegraphics[width=0.6\linewidth,keepaspectratio]{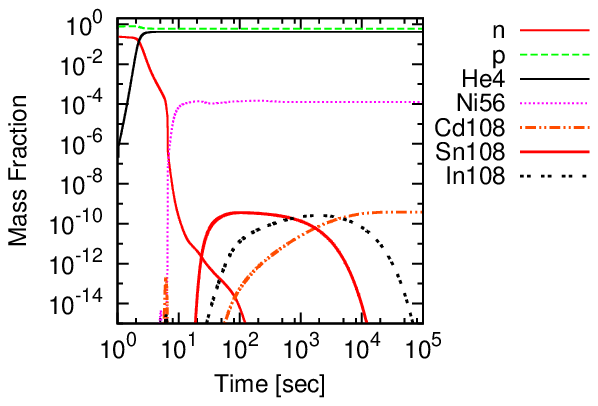}

 {(b) $\eta^{}_{high}=1.06\times10^{-3}_{}$}
\end{center}
\caption{Time evolution of the mass fractions in high-density regions of
  (a) $\eta^{}_{high}=1.02\times10^{-4}_{}$ and (b) $\eta^{}_{high}=1.06\times10^{-3}_{}$. }
\label{fig:MFNHSfig4}
\end{figure}

Figure~\ref{fig:MFNHSfig4} shows the results of nucleosynthesis in the
 high-density regions with $\eta^{}_{high}\simeq 10^{-4}_{}$ and $10^{-3}$.
 In Figure~\ref{fig:MFNHSfig4}(a), we see the time evolution of the
 abundances of Gd and Eu for the mass number 159. 
First ${}^{159}_{}$Tb (stable $r$-element) is synthesized and later
 ${}^{159}_{}$Gd and ${}^{159}_{}$Eu are synthesized through the neutron
 captures. After $t=10^3_{}$ sec, ${}^{159}_{}$Eu decays to nuclei
 by way of ${}^{159}_{}$Eu $\rightarrow {}^{159}_{}$Gd$\rightarrow {}^{159}_{}$Tb,
where the half-life of ${}^{159}_{}$Eu and ${}^{159}_{}$Gd are $26.1$ min
 and $18.479$~h, respectively. 

 For $\eta^{}_{high}\simeq 10^{-3}$, 
the result is seen in Figure~\ref{fig:MFNHSfig4}(b). ${}^{108}_{}$Sn which is
 proton-rich nuclei is synthesized. After that, stable nuclei
 ${}^{108}_{}$Cd is synthesized by way of
 ${}^{108}_{}$Sn $\rightarrow {}^{108}_{}$In $\rightarrow {}^{108}_{}$Cd,
 where the half-life of  ${}^{108}_{}$Sn and ${}^{108}_{}$In are 
$10.3$ min and  $58.0$ min, respectively. These results are qualitatively the same as 
Matsuura et al.~\cite{Matsuura2005}. 

 In addition, we notice  the production of radioactive nuclei of ${}^{56}_{}$Ni
and ${}^{57}_{}$Co, where
${}^{56}_{}$Ni is produced at early times, just after the formation of ${}^4_{}$He.
Usually,  nuclei such as ${}^{56}_{}$Ni and ${}^{57}_{}$Co 
are produced in supernova explosions, which are assumed to be the events
after the first star formation (e.g. Ref.~\cite{Hashimoto1995}). In IBBN
model, however, this production can be found to occur at extremely high density
region of $\eta^{}_{high}\geq 10^{-3}_{}$ as the primary elements
without supernova events in the early universe.

Final results ($T=4\times10^{7}$~K) of nucleosynthesis calculations are shown in
Table.~\ref{tab:abundance_4463_2}.
When we calculate the average values, we set the
abundances of $A>16$ to be zero for low-density side.
For $\eta^{}_{high}\simeq 10^{-4}_{}$, 
a lot of nuclei of $A>7$ are synthesized whose amounts are comparable
to that of ${}^{7}_{}$Li. Produced elements in this case include both
$s$-element (i.e., ${}^{138}_{}$Ba)
and $r$-elements (for instance, ${}^{142}$Ce and ${}^{148}$Nd).
For $\eta^{}_{high}\simeq 10^{-3}_{}$, there are few $r$-elements while both
$s$-elements (i.e., ${}^{82}_{}$Kr and ${}^{89}_{}$Y) and $p$-elements (i.e., ${}^{74}$Se and ${}^{78}$Kr) are synthesized such as the case of supernova explosions.
For $\eta^{}_{high} = 10^{-3}$ the heavy elements are produced slightly more than the total mass fraction (shown in Figure~\ref{fig:cntr_upLi7}) derived from the BBN code calculations.
This is because our BBN code used in \S \ref{sec:Results} includes the elements up to
$A=16$ and the actual abundance flow proceeds to much heavier elements.

\begin{figure}[t]

\begin{center}
 \includegraphics[width=.70\linewidth,keepaspectratio]{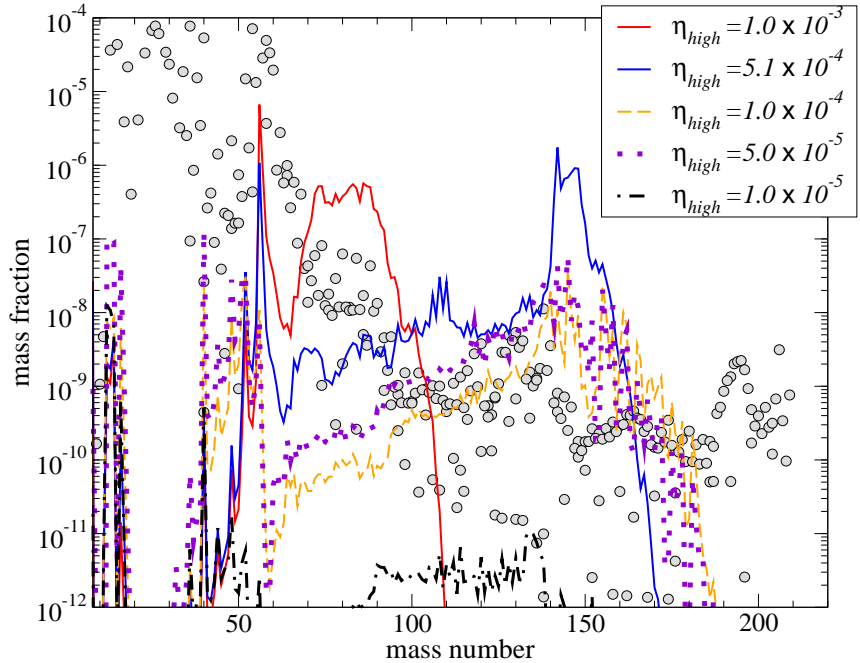}
\end{center}
\caption{Comparison of the averaged mass fractions in the two-zone model with the
 solar system abundances~\cite{Anders1989} (indicated by dots).}
\label{fig:massfrac_obs}
\end{figure}

Figure~\ref{fig:massfrac_obs} shows the abundances 
averaged between high- and low-density region using eq.~\eqref{eq:Yi_ave}
comparing with the solar system abundances~\cite{Anders1989}.
For $\eta^{}_{high}\simeq 10^{-4}_{}$, abundance productions of $120<A<180$ are
comparable to the solar values. For $\eta^{}_{high}\simeq10^{-3}_{}$,
those of $50<A<100$ have been synthesized well.
In the case of  $\eta^{}_{high}=5\times10^{-4}$, 
there are outstanding two peaks; one is around $A=56~(N=28)$ and the other can
be found around $A=140$.
Abundance patterns are very different from that of the solar system ones,
because IBBN  occurs under the condition of significant amount of abundances of
both neutrons and protons.
%---------------------------------------------------------------------------

\begin{table*}[tb]
\caption{\label{tab:abundance_4463_2}Mass fractions of heavy elements
 $(A~>~7)$ for three cases of
 $\eta^{}_{high}\simeq10^{-3}_{}, \eta_{high}=5.33\times10^{-4}$, 
and $\eta^{}_{high}\simeq10^{-4}$.}
\begin{center}
{\scriptsize
\begin{tabular}{ccc|ccc|ccc}
\hline\hline
%$~f^{}_v, R~$
\multicolumn{3}{c|}{$f^{}_v=3.23\times10^{-8}_{}, R=1.74\times10^{6}_{}$}
& \multicolumn{3}{c}{$f^{}_v=1.03\times10^{-7}_{}, R=9.00\times10^{5}_{}$}
& \multicolumn{3}{|c}{$f^{}_v=5.41\times10^{-7}_{}, R=1.84\times10^{5}_{}$}
 \\
%\hline
%$~\eta^{}_{high}$, $\eta^{}_{low}~$
 \multicolumn{3}{c|}{$\left( \eta^{}_{high}=1.06\times10^{-3}_{} \right)$}
& \multicolumn{3}{c}{$\left( \eta^{}_{high}=5.33\times10^{-4}_{} \right)$}
& \multicolumn{3}{|c}{$\left( \eta^{}_{high}=1.02\times10^{-4}_{} \right)$}
 \\
\hline
%$~t_{fin}, T_{fin}~$ &   
%\multicolumn{3}{c|}{~$t_{fin}=1.0\times10^{5}_{}$~sec,~$T_{fin}=4.2\times10^{7}_{}$~K~} &
%\multicolumn{3}{c|}{~$t_{fin}=1.1\times10^{5}_{}$~sec,~$T_{fin}=4.9\times10^{7}_{}$~K~} &
%\multicolumn{3}{c}{~$t_{fin}=1.2\times10^{5}_{}$~sec,~$T_{fin}=4.3\times10^{7}_{}$~K~} \\
\hline%\hline
element & high & average & 
element & high & average & 
element & high & average  \\
\hline\hline
Ni56 & $1.247\times 10^{-4}$ & $6.658\times 10^{-6}$ &  Nd142 & $2.051\times 10^{-5}$ & $1.738\times 10^{-6}$ &  Nd145 & $3.692\times 10^{-7}$ & $3.342\times 10^{-8}$ \\
Co57 & $1.590\times 10^{-5}$ & $8.487\times 10^{-7}$ &  Ni56 & $1.270\times 10^{-5}$ & $1.077\times 10^{-6}$ &  Ca40 & $2.706\times 10^{-7}$ & $2.450\times 10^{-8}$ \\
Sr86 & $1.061\times 10^{-5}$ & $5.662\times 10^{-7}$ &  Sm148 & $1.059\times 10^{-5}$ & $8.976\times 10^{-7}$ &  Mn52 & $2.417\times 10^{-7}$ & $2.188\times 10^{-8}$ \\
Sr87 & $9.772\times 10^{-6}$ & $5.214\times 10^{-7}$ &  Pm147 & $6.996\times 10^{-6}$ & $5.930\times 10^{-7}$ &  Eu155 & $2.374\times 10^{-7}$ & $2.149\times 10^{-8}$ \\
Se74 & $9.745\times 10^{-6}$ & $5.200\times 10^{-7}$ &  Pm145 & $6.559\times 10^{-6}$ & $5.559\times 10^{-7}$ &  Ce140 & $1.931\times 10^{-7}$ & $1.748\times 10^{-8}$ \\
Sr84 & $9.172\times 10^{-6}$ & $4.894\times 10^{-7}$ &  Sm146 & $6.539\times 10^{-6}$ & $5.542\times 10^{-7}$ &  Cr51 & $1.546\times 10^{-7}$ & $1.400\times 10^{-8}$ \\
Kr82 & $8.910\times 10^{-6}$ & $4.754\times 10^{-7}$ &  Nd143 & $4.146\times 10^{-6}$ & $3.514\times 10^{-7}$ &  Ce142 & $1.114\times 10^{-7}$ & $1.008\times 10^{-8}$ \\
Kr81 & $7.797\times 10^{-6}$ & $4.160\times 10^{-7}$ &  Pr141 & $3.957\times 10^{-6}$ & $3.354\times 10^{-7}$ &  Ni56 & $1.100\times 10^{-7}$ & $9.964\times 10^{-9}$ \\
Ge72 & $7.674\times 10^{-6}$ & $4.095\times 10^{-7}$ &  Nd144 & $3.952\times 10^{-6}$ & $3.350\times 10^{-7}$ &  Nd146 & $1.049\times 10^{-7}$ & $9.501\times 10^{-9}$ \\
Kr78 & $7.602\times 10^{-6}$ & $4.057\times 10^{-7}$ &  Sm147 & $3.752\times 10^{-6}$ & $3.180\times 10^{-7}$ &  Eu156 & $9.436\times 10^{-8}$ & $8.542\times 10^{-9}$ \\
Kr80 & $7.063\times 10^{-6}$ & $3.769\times 10^{-7}$ &  Sm149 & $3.322\times 10^{-6}$ & $2.815\times 10^{-7}$ &  Nd148 & $9.361\times 10^{-8}$ & $8.474\times 10^{-9}$ \\
Kr83 & $6.252\times 10^{-6}$ & $3.336\times 10^{-7}$ &  Pm146 & $2.629\times 10^{-6}$ & $2.228\times 10^{-7}$ &  Fe52 & $8.974\times 10^{-8}$ & $8.124\times 10^{-9}$ \\
Ge73 & $6.144\times 10^{-6}$ & $3.278\times 10^{-7}$ &  Sm144 & $2.207\times 10^{-6}$ & $1.870\times 10^{-7}$ &  Tb161 & $8.956\times 10^{-8}$ & $8.108\times 10^{-9}$ \\
Se76 & $5.929\times 10^{-6}$ & $3.164\times 10^{-7}$ &  Sm150 & $1.683\times 10^{-6}$ & $1.426\times 10^{-7}$ &  La139 & $8.804\times 10^{-8}$ & $7.971\times 10^{-9}$ \\
Br79 & $5.904\times 10^{-6}$ & $3.150\times 10^{-7}$ &  Pm144 & $1.581\times 10^{-6}$ & $1.340\times 10^{-7}$ &  N14 & $8.736\times 10^{-8}$ & $7.909\times 10^{-9}$ \\
Se77 & $5.345\times 10^{-6}$ & $2.852\times 10^{-7}$ &  Pm143 & $1.575\times 10^{-6}$ & $1.335\times 10^{-7}$ &  Cr48 & $8.561\times 10^{-8}$ & $7.750\times 10^{-9}$ \\
Y89  & $4.759\times 10^{-6}$ & $2.539\times 10^{-7}$ &  Sm145 & $1.010\times 10^{-6}$ & $8.568\times 10^{-8}$ &  Ba138 & $7.955\times 10^{-8}$ & $7.202\times 10^{-9}$ \\
Zr90 & $4.412\times 10^{-6}$ & $2.354\times 10^{-7}$ &  Co57 & $8.643\times 10^{-7}$ & $7.326\times 10^{-8}$  &  C12 & $7.672\times 10^{-8}$ & $6.945\times 10^{-9}$ \\
Rb85 & $4.324\times 10^{-6}$ & $2.307\times 10^{-7}$ &  Eu153 & $5.563\times 10^{-7}$ & $4.715\times 10^{-8}$ &  Dy162 & $6.835\times 10^{-8}$ & $6.188\times 10^{-9}$ \\
Rb83 & $4.082\times 10^{-6}$ & $2.178\times 10^{-7}$ &  Ce140 & $4.944\times 10^{-7}$ & $4.191\times 10^{-8}$ &  C13 & $6.428\times 10^{-8}$ & $5.819\times 10^{-9}$ \\
Y88  & $3.845\times 10^{-6}$ & $2.052\times 10^{-7}$ &  Nd145 & $4.376\times 10^{-7}$ & $3.709\times 10^{-8}$ &  O16 & $6.301\times 10^{-8}$ & $5.704\times 10^{-9}$ \\
Zr88 & $3.546\times 10^{-6}$ & $1.892\times 10^{-7}$ &  Eu155 & $4.224\times 10^{-7}$ & $3.581\times 10^{-8}$ &  Gd158 & $5.845\times 10^{-8}$ & $5.292\times 10^{-9}$ \\
As73 & $3.519\times 10^{-6}$ & $1.878\times 10^{-7}$ &  Eu151 & $4.106\times 10^{-7}$ & $3.480\times 10^{-8}$ &  Cs137 & $5.559\times 10^{-8}$ & $5.033\times 10^{-9}$ \\
Ga71 & $3.388\times 10^{-6}$ & $1.808\times 10^{-7}$ &  Cr52  & $4.071\times 10^{-7}$ & $3.450\times 10^{-8}$  &  Nd147 & $3.962\times 10^{-8}$ & $3.587\times 10^{-9}$ \\
Se75 & $2.933\times 10^{-6}$ & $1.565\times 10^{-7}$ &  Cd108 & $3.596\times 10^{-7}$ & $3.048\times 10^{-8}$ &  Ho165 & $3.770\times 10^{-8}$ & $3.413\times 10^{-9}$ \\
Nb91 & $2.896\times 10^{-6}$ & $1.545\times 10^{-7}$ &  Gd156 & $3.368\times 10^{-7}$ & $2.854\times 10^{-8}$ &   Pr143 & $3.111\times 10^{-8}$ & $2.817\times 10^{-9}$ \\
As75 & $2.856\times 10^{-6}$ & $1.524\times 10^{-7}$ &  Cd110 & $3.103\times 10^{-7}$ & $2.630\times 10^{-8}$ & Ce141 & $2.998\times 10^{-8}$ & $2.714\times 10^{-9}$ \\
Mo92 & $2.442\times 10^{-6}$ & $1.303\times 10^{-7}$ &  Eu152 & $2.809\times 10^{-7}$ & $2.381\times 10^{-8}$ &  Gd160 & $2.950\times 10^{-8}$ & $2.670\times 10^{-9}$ \\
Ge70 & $2.318\times 10^{-6}$ & $1.237\times 10^{-7}$ &  Sm151 & $2.795\times 10^{-7}$ & $2.369\times 10^{-8}$ &  Xe136 & $2.771\times 10^{-8}$ & $2.509\times 10^{-9}$ \\
Sr88 & $2.021\times 10^{-6}$ & $1.078\times 10^{-7}$ &  Eu154 & $2.759\times 10^{-7}$ & $2.339\times 10^{-8}$ &  Xe134 & $2.238\times 10^{-8}$ & $2.026\times 10^{-9}$ \\
\hline
$\displaystyle\sum_{A>7}{X(A)}$ &   $3.010\times10^{-4}_{}$
  & $1.652\times10^{-5}_{}$ & %Total $(A>7)$ 
$\displaystyle\sum_{A>7}{X(A)}$ &   $1.062\times10^{-4}_{}$
  & $9.006\times10^{-6}_{}$ & %Total $(A>7)$ 
  $\displaystyle \sum_{A>7}{X(A)}$ & $3.850\times10^{-6}_{}$  & $3.485\times10^{-7}_{}$ \\
\hline\hline
\end{tabular}
}
\end{center}
\end{table*}

\section{Summary and Discussion}
\label{sec:summary}

\begin{figure}[t]
\begin{center}
 \includegraphics[width=.70\linewidth,keepaspectratio]{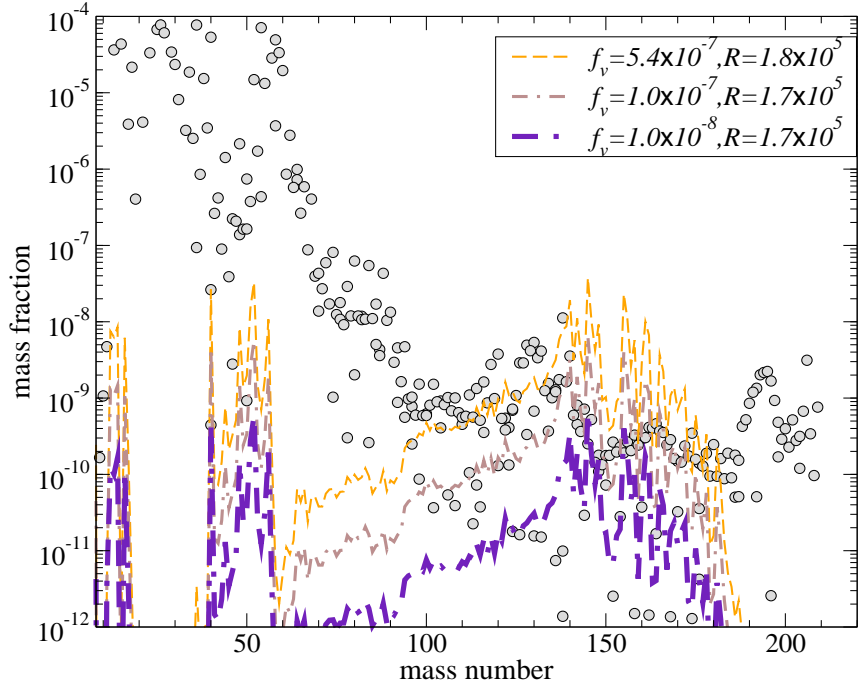}
\end{center}
\caption{Same as Fig.~\ref{fig:massfrac_obs}, but $\eta^{}_{high}$ is
fixed as $10^{-4}_{}$.}
\label{fig:massfrac_1e-4}
\end{figure}

We extend previous studies of 
Matsuura et al.~\cite{Matsuura2005,Matsuura2007} and investigate the consistency
between the light-element abundances in the IBBN model and
the observation of ${}^4$He and D/H.

First, we have done the nucleosynthesis calculation using the BBN code with 24 nuclei for 
the both regions.
The time evolution of the light-elements at the high-density region
differs significantly from 
that at the low-density region; 
The nucleosynthesis begins faster and
${}^4_{}$He is more abundant than that in the low density region.
By comparing the average abundances with ${}^4_{}$He and D/H observations, we can 
get the allowable parameters of the two-zone model: the volume fraction $f^{}_v$ of
the high-density region and the density ratio $R$ between the two regions.

Second, 
we calculate the nucleosynthesis that includes 4463 nuclei in the high-density
regions. Qualitatively, results of nucleosynthesis are the same as those in
Ref.~\cite{Matsuura2005}.
In the present results,
we showed that $p$- and $r$-elements are synthesized
simultaneously at high-density region with $\eta_{high}\simeq 10^{-4}$.
%Such a curious site of the nucleosynthesis has never been known in previous
%studies of nucleosynthesis.

We find that the average mass fractions in IBBN amount to
as much as the solar system abundances.
As see from Figure~\ref{fig:massfrac_obs}, there are over-produced elements
around $A=150$ (for $\eta^{}_{high}=10^{-4}$) and $A=80$ (for
$\eta_{high}=10^{-3}$).
Although it seems to conflict with the chemical evolution in the universe,
this problem could be solved by the careful choice of $f_v$ and/or $R$.
Figure \ref{fig:massfrac_1e-4} illustrates the mass fractions with
$\eta^{}_{high}=1.0\times10^{-4}_{}$ for three sets of $f_v-R$.
It is shown that the abundances can become lower than the solar
system abundances.
If we put constraint on the $f_v-R$ plane from the heavy element
observations~\cite{Christlieb2007,Frebel2007,Siqueira2013,Hill2013}, 
the parameters in IBBN model should be tightly determined.

In the meanwhile, we would like to touch on the consistency against the primordial
${}^7_{}$Li. 
We have obtained interesting results about ${}^7_{}$Li abundances in our model.
For the recent study of ${}^7_{}$Li, the lithium problem arises from the 
discrepancy among ${}^7_{}$Li abundance predicted by SBBN theory, 
the baryon density of WMAP, and abundance inferred from observations of 
metal-poor stars (see Refs.~\cite{Coc2004,Cyburt2008}). 
As seen in Table~\ref{tab:abundance_4463}, 
 ${}^7_{}$Li is clearly overproduced such as ${}^7_{}$Li/H$|^{}_{ave}=1.52\times10^{-9}$
for $\eta^{}_{high}=10^{-5}_{}$, although we adopt the highest
observational value ${}^7_{}$Li/H$=\left( 2.75 - 4.17 \right)\times10^{-10}$~\cite{Korn2006}.
However, for cases of $\eta^{}_{high}=10^{-3}, 5\times10^{-4}_{}$, and $10^{-4}$,
the values of ${}^7_{}$Li/H$|^{}_{ave}$ agree with the observation.
Usually, the consistency with BBN  has been checked using observations of ${}^4_{}$He, D/H, and
${}^7_{}$Li/H. Then the parameters such as $\eta^{}_{high}=10^{-5}_{}$
ought to be excluded. However the abundance of ${}^7_{}$Li/H$|^{}_{ave}$ is
sensitive to the values of both $\eta^{}_{high}$ and $\eta^{}_{low}$.
As the future work of IBBN, we will study in detail the  ${}^7_{}$Li production. 
In addition, recent ${}^4_{}$He observation could suggest the need
of non-standard BBN model~\cite{Izotov2010}. IBBN may also give a clue
 to the problems.

\begin{acknowledgement}
This work has been supported in part by a Grant-in-Aid for Scientific
Research (24540278) of the Ministry of Education,
Culture, Sports, Science and Technology of Japan, 
and in part by a grant for Basic Science Research Projects 
from the Sumitomo Foundation (No. 080933).
\end{acknowledgement}

%\appendix
%\section{First Appendix} %Empty argument \section{} yields `Appendix'. 
%
%\section{Second Appendix}


\begin{thebibliography}{99}
%%%%%%%%%%%%%%%%%%%%%%%%%%%%%%%%%%%%%%%%%%%%%%%%%%%%%%%%%%%%%
% Some macros are available for the bibliography:
%  o for general use
%    \JL : general journals                 \andvol : Vol (Year) Page
%  o for individual journal 
%    \AJ   : Astrophys. J.           \NC         : Nuovo Cim.
%    \ANN  : Ann. of Phys.           \NPA, \NPB  : Nucl. Phys. [A,B]
%    \CMP  : Commun. Math. Phys.     \PLA, \PLB  : Phys. Lett. [A,B]
%    \IJMP : Int. J. Mod. Phys.      \PRA - \PRE : Phys. Rev. [A-E]     
%    \JHEP : J. High Energy Phys.    \PRL        : Phys. Rev. Lett.
%    \JMP  : J. Math. Phys.          \PRP        : Phys. Rep.
%    \JP   : J. of Phys.             \PTP        : Prog. Theor. Phys.     
%    \JPSJ : J. Phys. Soc. Jpn.      \PTPS       : Prog. Theor. Phys. Suppl.
% Usage:
%  \PRD{45,1990,345}          ==> Phys.~Rev.\ \textbf{D45} (1990), 345
%  \JL{Nature,418,2002,123}   ==> Nature \textbf{418} (2002), 123
%  \andvol{B123,1995,1020}    ==> \textbf{B123} (1995), 1020
%%%%%%%%%%%%%%%%%%%%%%%%%%%%%%%%%%%%%%%%%%%%%%%%%%%%%%%%%%%%%
\bibitem{Matsuura2005}
	S.~Matsuura, S.~I.~Fujimoto, S.~Nishimura, M.~A.~Hashimoto and K.~Sato,
%	``Heavy Element Production in Inhomogeneous Big Bang Nucleosynthesis,''
	Phys.\ Rev.\  D {\bf 72}, 123505 (2005)
\bibitem{PDG2012}
	J.~Beringer {\it et al.}  [Particle Data Group],
	Phys.\ Rev.\  D {\bf 86}, 010001 (2012).
%\bibitem{PDG2010} 
%	K.~Nakamura, \& Particle Data Group 2010, 
%	Journal of Physics G Nuclear Physics, {\bf 37}, 075021, (2010)
\bibitem{Iocco:2008va}
	G.~Steigman,
%	``Primordial Nucleosynthesis in the Precision Cosmology Era,''
	Ann.\ Rev.\ Nucl.\ Part.\ Sci.\  {\bf 57}, 463 (2007); \\
	F.~Iocco, G.~Mangano, G.~Miele, O.~Pisanti and P.~D.~Serpico,
	Phys.\ Rept.\  {\bf 472}, 1 (2009)
\bibitem{Coc2012} 
	Coc, A., Goriely, S., Xu, 
	Y., Saimpert, M., \& Vangioni, E.\ 2012, \Apj, 744, 158 
 \bibitem{Luridiana2003}
	 V.~Luridiana,A.~Peimbert, M.~Peimbert, \&  M.~Cervino, 
	 Astrophys.\ J.\ {\bf 592}, 846 (2003)
\bibitem{OliveSkillman04}
	Olive \& Skillman, Astrophys. J., {\bf 617}, 29--40, (2004)
% \bibitem{Izotov:2007ed}
%  Y.~I.~Izotov, T.~X.~Thuan and G.~Stasinska,
%  ``The primordial abundance of 4He: a self-consistent empirical analysis of
%  systematic effects in a large sample of low-metallicity HII regions,''
  Astrophys.\ J.\  {\bf 662}, 15 (2007)
\bibitem{Kirkman2003}
  D.~Kirkman, D.~Tytler, N.~Suzuki, J.~M.~O'Meara and D.~Lubin,
%   ``The cosmological baryon density from the deuterium to hydrogen ratio
%  towards QSO absorption systems: D/H towards Q1243+3047,''
  Astrophys.\ J.\ Suppl.\  {\bf 149}, 1 (2003)
  [arXiv:astro-ph/0302006].
  %%CITATION = ASTRO-PH 0302006;%%
 \bibitem{OMeara2006}
	 J.~M.~O'Meara, S.~Burles, J.~X.~Prochaska, G.~E.~Prochter, R.~A.~Bernstein and K.~M.~Burgess,
%	 ``The Deuterium to Hydrogen Abundance Ratio Towards the QSO SDSS1558-0031,''
	 Astrophys.\ J.\  {\bf 649}, L61 (2006)
\bibitem{Pettini2008} 
	M.~Pettini, B.~J. Zych, M.~T. Murphy, A. Lewis,
	\& C. C. Steidel, Mon. Not. R. Astron. Soc. {\bf 391}, 1499, (2008)
\bibitem{Bania2002}
	T.~M.~Bania, R.~T.~Rood and D.~S.~Balser,
%	``The cosmological density of baryons from observations of
	$^3$He$^+$ in the Milky Way,''
	Nature {\bf 415}, 54 (2002).
\bibitem{2003ApJ...585..611V} Vangioni-Flam, 
E., Olive, K.~A., Fields, B.~D., \& Cass{\'e}, M.\ 2003, \Apj, 585, 611 
\bibitem{WMAP5}
	C.~L.~Bennett, et al., {\it et al.}, arXive:1212.5225 [astro-ph.CO]
\bibitem{Matsuura:2004ss}
	S.~Matsuura, A.~D.~Dolgov, S.~Nagataki and K.~Sato,
%	``Affleck-Dine Baryogenesis and heavy elements production from Inhomogeneous
%	Big Bang Nucleosynthesis,''
	Prog.\ Theor.\ Phys.\  {\bf 112}, 971 (2004)
\bibitem{Alcock1987}
	C.~Alcock, G.M.~Fuller, and G.J.~Mathews, Astrophys. J. {\bf 320}, 439 (1987)
\bibitem{Fuller1988}
	G.~M.~Fuller, G.~J.~Mathews and C.~R.~Alcock, Phys.\ Rev.\  D {\bf 37}, 1380 (1988);
\bibitem{IBBN_QCD}
	H.~Kurki-Suonio and R.~A.~Matzner, Phys.Rev. {\bf D39}, 1046 (1989);  \\
	H.~Kurki-Suonio and R.~A.~Matzner, Phys.Rev. {\bf D42}, 1047 (1990); 
\bibitem{WMAP3}
	C.~L.~Bennett, {\it et al.}, \Apj Suppl. {\bf 148}, 1 (2003) \\
	D.~N.~Spergel {\it et al.},  Astrophys.\ J.\ Suppl.\  {\bf 170},
	377 (2007) \\
	J.~Dunkley {\it et al.}   Astrophys.\ J.\ Suppl.\  {\bf 180}, 306 (2009) 
\bibitem{IBBN0}
	J. H. Applegate, C. J. Hogan, and R. J. Scherrer, Phys. Rev. {\bf D35}, 1151 (1987)
\bibitem{IBBN1}
	R. M.~Malaney and W. A.~Fowler, Astrophys. J {\bf 333}, 14 (1988);\\ 
	J. H. Applegate, C. J. Hogan, R. J. Scherrer, Astrophys. J. {\bf 329}, 572 (1988);\\
	N. Terasawa and K. Sato, Astrophys. J. {\bf 362}, L.47 (1990);\\
	D.~Thomas, D.~N.~Schramm, K.A.~Olive, G.~J.~Mathews, B.~S.~Meyer, and B.~D.~
	Fields, \Apj {\bf430}, 291 (1994);
 \bibitem{TerasawaSato89}
	 N.~Terasawa and K.~Sato,
	 Phys.\ Rev.\  D {\bf 39}, 2893 (1989)
\bibitem{2zone}
	K.~Jedamzik, and J.B.~Rehm, Phys. Rev. {\bf D64}, 023510 (2001)[astro-ph/0101292];\\
	T.~Rauscher, H.~Applegate, J.~Cowan, F.~Thielmann, and M.~Wiescher, \Apj {\bf 429},
	499 (1994).
 \bibitem{Jedamzik1994}
	K.~Jedamzik, G.~M.~Fuller, G.~J.~Mathews, and T.~Kajino, \Apj {\bf 422}, 423 (1994);
\bibitem{Wagoner1967}
	R.~V. Wagoner, W.~A. Fowler,  \& F. Hoyle, 
	Astrophys.\  J.\ , {\bf 148}, 3 (1967)
 \bibitem{Wagoner1972}
	 R.~V.~Wagoner,
	 ``Big Bang Nucleosynthesis Revisited,''
	 Astrophys.\ J.\  {\bf 179}, 343 (1973).
	 %%CITATION = ASJOA,179,343;%%
\bibitem{Juarez2009} 
	Y. Juarez, R. Maiolino, R. Mujica, M. Pedani, 
	S. Marinoni, T. Nagao, A. Marconi, \& E. Oliva, Astron. \& Astrophys.,
	494, L25, (2009)
\bibitem{Moriya2010}
  T.~Moriya and T.~Shigeyama,
%  ``Multiple Main Sequence of Globular Clusters as a Result of Inhomogeneous
%  Big Bang Nucleosynthesis,''
  Phys.\ Rev.\  D {\bf 81}, 043004 (2010)
\bibitem{Bedin2004}
	L. R. Bedin et al., Astrophys. J., {\bf 605}, L125 (2004); 
\bibitem{Piotto2007}
	G. Piotto et al., Astrophys. J., {\bf 661} L53, (2007)
  %\cite{affl-dine}
 \bibitem{affl-dine}
	 I. Affleck, and M. Dine, Nucl. Phys. {\bf B249}, 361 (1985).
\bibitem{Rauther2006}
	T.~Rauscher, Phys.\ Rev.\  D {\bf 75}, 068301 (2007)
\bibitem{Matsuura2007}
	S.~Matsuura, S.~I.~Fujimoto, M.~A.~Hashimoto and K.~Sato,
%	``Reply to 'Comment on 'Heavy element production in inhomogeneous big bang
	nucleosynthesis'',''
	Phys.\ Rev.\  D {\bf 75}, 068302 (2007).
\bibitem{Hashimoto1985}
	M. Hashimoto \& K. Arai, Physics Reports of Kumamoto University,
	{\bf 7}, 47, (1985).
\bibitem{DAA04}
	P.~Descouvemont, A.~Adahchour, C.~Angulo, A.~Coc,
	\& E.~Vangioni-Flam,
	Atomic Data and Nuclear Data Tables, 88, 203 (2004)
%\bibitem{NACRE}
%	C.~Angulo, M.~Arnould, M.~Rayet, P.~Descouvemont, D.~Baye,
%	C.~ Leclercq-Willain, A.~Coc, S.~Barhoumi, P.~Aguer,
%	C.~Rolfs, et al., Nuclear Physics A 656, 3 (1999).
\bibitem{Izotov2010}
	Y. Izotov and T. X. Thuan, \Apj, {\bf 710}, L67  (2010)
%\bibitem{Hagiwara:2002fs}
%	K.~Hagiwara {\it et al.}  [Particle Data Group],
 % ``Review of particle physics,''
%	Phys.\ Rev.\  D {\bf 66}, 010001 (2002). 
\bibitem{Aver2012}
	E.~Aver, K.~A.~Olive, \& E.~D.~Skillman, JCAP, {\bf 04}, 004 (2012)	
%\bibitem{PDG2008}
%	B.~Fields %and S.~Sarkar,
%	``Big-bang nucleosynthesis (PDG mini-review),''
%	arXiv:astro-ph/0601514.
\bibitem{Pettini2012} 
	M.~Pettini,\& R.~Cooke, Mon. Not. R. Astron. Soc. {\bf 425}, 2447, (2012)
%\bibitem{jedam}
%	 K.~Jedamzik [astro-ph/9911242].
\bibitem{fujimoto} 
	S. Fujimoto,M. Hashimoto, O. Koike,K. Arai, \& R. Matsuba, \Apj {\bf
	585}, 418 (2003),\\
	O. Koike, M. Hashimoto, R. Kuromizu, \& S. Fujimoto, \Apj {\bf603}, 592 (2004),\\
	S. Fujimoto, M. Hashimoto, K. Arai, \& R. Matsuba, \Apj,
	{\bf 614}, 847 (2004),\\
	S.~Nishimura, K.~Kotake, M.~Hashimoto, S.~Yamada, N.~Nishimura,
	S.~Fujimoto and K.~Sato, Astrophys.\ J.\  {\bf 642}, 410 (2006).
\bibitem{Kawano}
	L. Kawano, FERMILAB-Pub-92/04-A  
\bibitem{Anders1989}
	E.~Anders and N.~Grevesse,
%  ``Abundances Of The Elements: Meteroritic And Solar,''
	Geochim.\ Cosmochim.\ Acta {\bf 53}, 197 (1989).
\bibitem{Hashimoto1995} 
	M. Hashimoto, Progress of Theoretical Physics, 94, 663, (1995).
\bibitem{Anderson2009}
  M.~E.~Anderson, J.~N.~Bregman, S.~C.~Butler and C.~R.~Mullis,
  Astrophys.\ J.\  {\bf 698}, 317 (2009)
\bibitem{Coc2004}
  A.~Coc, E.~Vangioni-Flam, P.~Descouvemont, A.~Adahchour and C.~Angulo,
%  ``Updated Big Bang Nucleosynthesis confronted to WMAP observations and to the
%  Abundance of Light Elements,''
  Astrophys.\ J.\  {\bf 600}, 544 (2004)
\bibitem{Cyburt2008}
  R.~H.~Cyburt, B.~D.~Fields and K.~A.~Olive,
%  ``A Bitter Pill: The Primordial Lithium Problem Worsens,''
  JCAP {\bf 0811} 012, (2008)
\bibitem{Korn2006}
  A.~J.~Korn {\it et al.},
%  ``A probable stellar solution to the cosmological lithium discrepancy,''
  Nature {\bf 442} (2006) 657
  [arXiv:astro-ph/0608201].
%%
\bibitem{Christlieb2007} 
Frebel, A., Christlieb, N., Norris, J.~E., et al.\ 2007, \Apj, 660, L117 
\bibitem{Frebel2007} Frebel, A., Norris, 
J.~E., Aoki, W., et al.\ 2007, \Apj, 658, 534 
\bibitem{Siqueira2013} 
Siqueira Mello, C., Spite, M., Barbuy, B., et al.\ 2013, \aap, 550, A122 
\bibitem{Hill2013} 
Worley, C.~C., Hill, V., Sobeck, J., \& Carretta, E.\ 2013, \aap, 553, A47 
\end{thebibliography}
\end{document}